\documentclass[useAMS,usenatbib]{mn2e}
\usepackage{epsfig}
\usepackage{natbib}
\usepackage{lscape}
\usepackage{subfigure}
\usepackage{verbatim}
\usepackage{amssymb,amsmath}
\DeclareMathVersion{bold}

\title[VLBI Search for Double Black Holes]{A Radio Census of Binary Supermassive Black Holes}

\author[S. Burke-Spolaor]{ 
S. Burke-Spolaor$^{1,2}$\thanks{Email: sburke@astro.swin.edu.au}\\
$^{1}$Swinburne University of Technology Centre for Astrophysics and Supercomputing, Hawthorn VIC, Australia\\
$^{2}$Australia Telescope National Facility, CSIRO, P.O. Box 76, Epping, NSW 1710, Australia\\
}
\date{}
\begin{document}

\maketitle
\begin{abstract}
  Using archival VLBI data for 3114 radio-luminous active galactic nuclei, we
  searched for binary supermassive black holes using a radio spectral index
  mapping technique which targets spatially resolved, double radio-emitting
  nuclei. Only one source was detected as a double nucleus. This result is
  compared with a cosmological merger rate model and interpreted in terms of
  (1) implications for post-merger timescales for centralisation of the two
  black holes, (2) implications for the possibility of ``stalled'' systems,
  and (3) the relationship of radio activity in nuclei to mergers. Our
  analysis suggests that binary pair evolution of supermassive black holes
  (both of masses $\geq 10^8\,{\rm M}_\odot$) spends less than 500\,Myr in
  progression from the merging of galactic stellar cores to within the
  purported stalling radius for supermassive black hole pairs. The data show
  no evidence for an excess of stalled binary systems at small separations. We
  see circumstantial evidence that the relative state of radio emission
  between paired supermassive black holes is correlated within orbital
  separations of 2.5\,kpc.
\end{abstract}
\begin{keywords}
\end{keywords}


\section{Introduction}\label{sec:intro}
Models of the Universe in which hierarchical merging dominates the growth of
galaxies have strong predictions for the presence of binary supermassive black
holes (SMBHs) at galaxy centres \citep[e.g.][]{VHM}. As galaxies containing
such massive black holes collide, the central black holes are expected to
inspiral and form a bound interacting system which will have a significant
impact on the central galactic environment \citep{merritt06}. In the time
preceding their coalescence, these binary sources will be among the brightest
of gravitational radiation sources detectable in the very low frequency (nHz
to $\mu$Hz) spectrum by pulsar timing experiments \citep[e.g.][]{jenetPPTA}.
Such experiments are sensitive to both individual binary SMBHs with sub-parsec
orbits and the stochastic gravitational wave background made up of the
collective signal from binaries of mass $\sim$$10^6 - 10^9$\,M$_\odot$
\citep{detweiler79}. In the mid-frequency spectrum ($10^{-4}$\,Hz to 0.1\,Hz),
the space-based Laser Interferometer Space Antenna (LISA)\footnote{See
  http://lisa.nasa.gov}, is sensitive to gravitational waves from individual
binary SMBH sytems of mass $\lesssim 10^7$\,M$_\odot$ in the later stages of
their inspiral, coalescence, and post-coalescence ringdown. Predictions of the
strength of this astrophysical gravitational wave background and the expected
detection rates for LISA and pulsar timing of individual black holes are based
on parameterisations of merger rates and the SMBH population
\citep{rajogopalromani,jaffebacker,wyitheloeb,enoki,sesanagwb,sesanasingle},
giving these gravitational wave detectors the potential to provide unique
insights into the processes of galaxy formation.

In theoretical treatments, however, the evolution of post-merger central
supermassive black holes still holds significant uncertainties. The
steps in binary black hole formation and evolution were first laid out
by \citet{begelmanetal80}. After a galaxy pair becomes virially bound,
dynamical friction and violent relaxation drive the stellar cores
containing the massive black holes (of masses $m_1$ and $m_2;
m_1>m_2$) to the centre of the merger remnant on roughly a dynamical
friction time. 
%
When the cores have merged, the black holes independently experience
dynamical friction against the merged core's stellar environment. This
further centralises each black hole, leading the system to become a
bound pair at a separation where the enclosed stellar mass equals the
total binary mass. If the stellar core is modelled as a single
isothermal sphere with stellar density $\rho_{*}(r)=\sigma_v^2/(2\pi
Gr^2$), this occurs when the black holes reach a orbital semi-major
axis of
\begin{equation} \label{eq:abin}
	a_{\rm{bin}}=\frac{3}{2}\frac{G(m_1+m_2)}{\sigma_v^2}~,
\end{equation}
where $\sigma_v$ is the velocity dispersion of the merger remnant. The
length of time spent by black holes to reach this stage follows the
Chandrasekhar timescale for dynamical friction, and is limited by the
inspiral timescale for the less massive black hole
\citep[cf.][]{laceycole}:
\begin{equation}\label{eq:tdf}
	t_{\rm{df}} = 1.654 \frac{r^2\sigma_v}{Gm_2\rm{ln}\Lambda}~,
\end{equation}
where we have assumed circular black hole orbits, and $r$ is the
orbital radius of the less massive black hole from the centre of the
galactic potential. In such a system, we can assume a value for the
Coulomb logarithm $\rm{ln}\Lambda\simeq 5$ (a non-circular orbit with
ellipticity $\epsilon=0.5$ will add a factor of roughly 0.5 to the
timescale and the logarithm will be $2\lesssim\rm{ln}\Lambda\lesssim
3$; e.g. \citealt{gualandrismerritt08}). The black holes will form a
binary when $r=a_{\rm{bin}}$.

The orbital evolution of the black holes proceeds further due to dynamical
friction, however as the binary tightens and increases in velocity, dynamical
friction becomes an ineffective means of energy transfer and the rate of
inspiral slows. In this intermediate inspiral stage, there are significant
uncertainties in both inspiral mechanism and timescale. Relaxing stars in
radial orbits entering the ``loss cone'' of the massive binary
\citep{frankrees76} will undergo 3-body interactions with the SMBH binary
system and be ejected, carrying away angular momentum and further shrinking
the binary orbit. However, without an additional mechanism for energy transfer
or a means to efficiently refill the loss cone, the binary inspiral halts as
the loss cone is emptied. Recognising the long timescale that loss cone
re-population might take for binary systems, \citet{begelmanetal80} suggest
the possibility of gas ejection from the system or gas accretion onto the
larger black hole, which will cause shrinkage in the binary orbit to conserve
angular momentum \citep[see also, e.\,g.,][]{lreview}. If some intermediate
process is able to sufficiently shrink the binary orbit, gravitational
radiation will cause a binary with ellipticity $\epsilon=0$ to coalesce in a
timescale
\begin{equation}\label{eq:tgrav}
	t_g = \frac{5c^5}{256G^3}\frac{a^4}{m_1m_2(m_1+m_2)}~.
\end{equation}

N-body simulations and semi-analytical models have manifested various
intermediate inspiral processes (see, e.g., the extensive review of
\citealt{bigreview}). The ``last parsec problem'' is the name given to the
hurdle encountered by nearly all merger models in which the intermediate
inspiral mechanisms have timescales sufficiently long that they are unable to
bring the binary to a regime in which the emission of gravitational radiation
can drive the black holes to coalescence in less than a Hubble time. If this
is the case, there may be many instances of ``stalled'' binary objects which
spend a large portion of their lifetimes with orbital radii within the range
0.01-10\,pc \citep{yu02}. A functional form for the stalling radius is
estimated by \citet{merritt06}:
\begin{equation}\label{eq:astall}
	\frac{a_{\rm stall}}{2~a_{\rm bin}} = 0.2~\frac{m_1/m_2}{(1+m_1/m_2)^2}~.
\end{equation}
If systems are unable to find the fuel to reach coalescence in less
than the age of the Universe, then they may appear to stall
indefinitely at orbits greater (thus GW frequencies lower, and
amplitudes generally smaller) than those in the range detectable by
either pulsar timing arrays or LISA. This would have dramatic
consequences for GW detectors and for hierarchical formation models
alike.

Since few small-orbit binary systems are known and electromagnetic
tracers of post-merger galaxy cores are hard to identify, it has been
difficult to study the post-merger dynamics of binary black hole
systems. With only two confirmed sub-kpc double black holes
serendipitously identified, (NGC 6240 at $\sim$1\,kpc,
\citealt{ngc6240}; 0402+379 at 7\,pc, \citealt{rodriguezetal06}) there
is little possibility to assess the physical mechanism responsible for
driving binary systems into the phase when gravitational radiation
dominates the binary inspiral. Observationally, there are several ways
to identify galaxies with binary nuclei, which are reviewed briefly below
(see also \citealt{komossa06} for a good overview).

Pairs of galaxies/quasars with small angular separations and similar
redshifts, as well as galaxies which show disturbed structure and tidal tails
in the optical and infrared, can indicate interaction between widely separated
stellar cores containing SMBHs (e.g.
\citealt{myersetal08,deproprisetal07,johnny}). Secondary signals that have
been modelled as a consequence of a close or bound binary SMBH include
quasi-periodic oscillations in flux density (OJ287, \citealt{valtonenetal88}),
abnormal jet morphology such as helices or X-shaped radio jets
\citep{saripallietal08,lalandfriends}, spatial oscillations of an active
galactic nucleus (3C66B, \citealt{sudouetal03}), and sources with double or
offset broad/narrow line regions, implying active galactic nuclei (AGN) moving
at high relative velocities to each other or to the host galaxy (e.g.
\citealt{petersonetal87,comerfordetal08,nature09,decarli}). In most of these cases,
other explanations have better suited the observed characteristics of each
source. However, spatially resolved systems such as 3C75, which shows two
jet-emitting radio AGN cores, and NGC 6240 \citep{ngc6240} and other double
X-ray emitting AGN \citep[e.g.][]{dickref}, are straightforward to identify
and can provide a direct detection of a double SMBH system.

The core regions of radio-luminous AGN have been observed to show a distinct
radio continuum spectrum which typically peaks at radio frequencies above
$\sim$1\,GHz, giving core components a characteristically flat two-point
spectral index ($\alpha> -0.5$, where radio flux $F_\nu\propto \nu^\alpha$) at
GHz frequencies \citep[e.g.][]{slee94}. This feature distinguishes the region
most directly associated with the host black hole from the outer features such
as larger scale jets or hotspots, which show power-law spectra generally at or
steeper than $\alpha=-0.7$. Taking advantage of these unique spectral
signatures of the core regions of radio-emitting AGN at high radio frequency
(1 to 20\,GHz) and the sub-milliarcsecond precision of observations using
radio Very Long Baseline Interferometry (VLBI), it is possible to spatially
resolve and identify AGN containing multiple cores using multifrequency
gigahertz spectral imaging. This property has also been utilized to locate
gravitationally lensed compact radio sources; if an AGN is lensed, it will
also appear as a spatially resolved pair with equivalent radio continuum
spectra (note that there is direct overlap in the target of such such
searches; because e.\,g. \citealt{wilk01} found no lenses in 300 VLBI images,
this implies no SMBH binaries were found, either).

In the case of a genuine physical pair, resolutions of 1\,milliarcsecond are
able to discern components with a projected spacing of above $\sim$8.5\,pc at
\emph{all} redshifts, and down to sub-pc resolutions for the nearest
galaxies. This makes VLBI observations well-suited to explore double
supermassive black hole systems even at very small projected separations, and
within post-merger galaxies that are otherwise unidentifiable as double or
disturbed systems. Identification of double black holes in post-merger systems
and binary black holes in the intermediate stage of inspiral will aid in
showing evidence for stalling and may provide statistical estimates of the
rate of SMBH inspiral at various phases.


This paper reports on the first systematic search of a large number of radio
sources for spatially resolved binary radio AGN using archival VLBI data. We
present a brief description of the archival data sample in \S\ref{sec:sample},
and summarise our search technique in \S\ref{sec:technique}. The candidates
resulting from this search are assessed in \S\ref{sec:results}. We outline an
interpretive framework for making statistical estimates of inspiral timescales
based on our detections in \S\ref{sec:interp}, and give the results of this
analysis in \S\ref{sec:timeresults}. In \S\ref{sec:discussion} we discuss the
analysis in terms of implications for black hole inspiral rates, stalling,
radio core emission in the merging process, and implications for the
pulsar-timing-detectable gravitational wave signal from SMBH binaries.

\begin{figure}
{
 \includegraphics[width=0.48\textwidth]{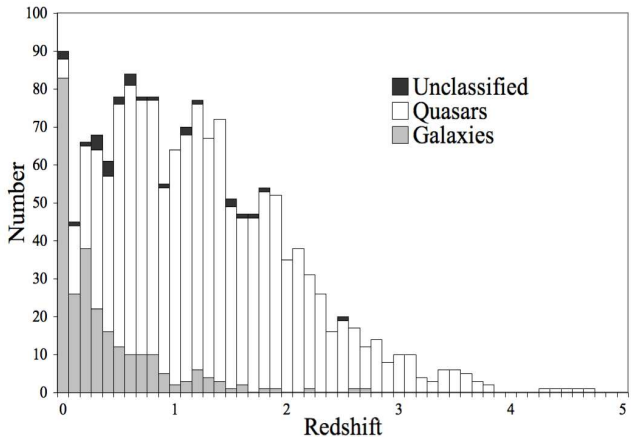}
}
\caption{The redshift distribution of 1575 objects in our sample. The closest source is at $z=0.000113$, while the highest redshift source is a quasar at $z=4.715$.}
\label{fig:redshifts}
\end{figure}

\section{The Archival Data Sample}\label{sec:sample}
The data used in this search were extracted from the VLBI archives maintained
online as the Goddard Space Flight Center astrometric and geodetic
catalogues.\footnote{\mbox{http://lacerta.gsfc.nasa.gov/vlbi/solutions/} ---
Please see website and references therein for a complete description of
original data sets} The catalogue version used was ``VLBI global solution
2008a\_astro,'' which contains a large volume of calibrated VLBI data from
1980 to early 2008, including data from the VLBA calibrator surveys
\citep{VCS1,VCS2,VCS3,VCS4,VCS5,VCS6}. The catalogue contained observations
unevenly sampled across 2.2\,GHz to 43\,GHz for 4169 radio sources. For our
search technique, we are limited to sources observed at two or more
frequencies, and to avoid potentially problematic intrinsic source variability
(see \S \ref{sec:technique}), we rejected frequency pairs which were observed
at times separated by more than 35 days. Our sample was further culled because
of software incompatibility for some data, and for the remaining dataset,
observations were rejected if the images revealed data which contained no
clear source above the noise. Three sources were removed because a literature
searched revealed them to be Galactic HII regions or stars.  In total, we were
able to search data for 3114 sources.

Approximately half of the sources in our sample had redshift information
available. The distribution for the 1575 sources of known redshift is shown in
Fig. \ref{fig:redshifts}; the source types and redshifts were gathered from
the NASA Extragalactic Database.\footnote{NED,
\mbox{http://nedwww.ipac.caltech.edu/}} The mean redshifts of our sample are
$z_{g}=0.402, z_{q}=1.401,$ and $z_{tot}=1.226$ for galaxies, quasars, and all
the sources with known redshift, respectively.
\begin{table}
  \centering
    \begin{tabular}{| r | c | c c c c }
      & \textbf{2} & \textbf{5} & \textbf{8} & \textbf{15-24} & \textbf{43} \\
        \textbf{5}  & 96     & ---  &          &          & \\
        \textbf{8}  & 3101& 94  & ---     &          & \\
\textbf{15-24}  & 190  & 59  & 239  & ---       & \\
        \textbf{43}& 34    & 0     & 37    & 132   & ---\\
    \end{tabular}
\caption{Frequency band pairs usable in our search; all frequencies are in units of gigahertz. The frequencies above refer to the rough observing band only.
In total there were 3982 frequency pairs for 3114 sources.}\label{table:nsources}
\end{table}

\section{VLBI search for double AGN}\label{sec:technique}
Our technique for identifying binary active galactic nuclei using VLBI is
essentially a search for objects which show more than one flat-spectrum
component. It requires morphological analysis and spectral index imaging
across two or more frequencies to distinguish the nucleus of an object from
its other components, such as complex extended structure or bright, unresolved
hotspots along a jet. These other source components, as was previously noted,
show systematically declining synchrotron spectra at $\alpha \lesssim
-0.7$. The technique and the use of archival VLBI data require several
precautions, primarily because both the accuracy of frequency-dependent flux
density measurements and morphological fidelity of the radio image are
important to this method.

\begin{figure*}
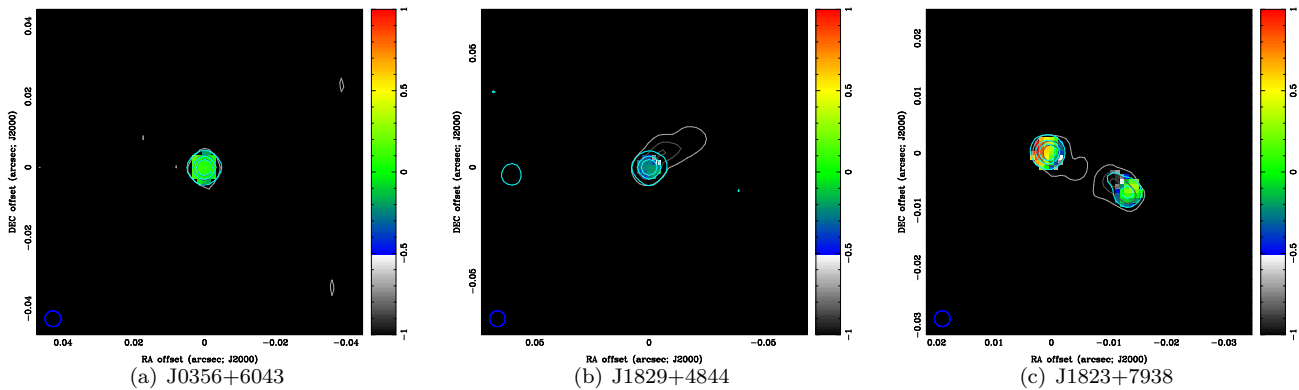

 \subfigure[J0356+6043]
{
 \includegraphics[height=0.3\textwidth,angle=270]{ps/J0356+6043SX.ps}
}
\quad
 \subfigure[J1829+4844]
{
 \includegraphics[height=0.3\textwidth,angle=270]{ps/J1829+4844SX.ps}
}
\quad
 \subfigure[J1823+7938]
{
 \includegraphics[height=0.3\textwidth,angle=270]{ps/J1823+7938SX.align.ps}
}
\caption{Two-frequency spectral index maps for sources tagged as \emph{(a)}
  pointlike, \emph{(b)} resolved, and \emph{(c)} multiple flat-spectrum
  component sources. All maps show matched-resolution data. The color scale
  represents the spectral index calculated from two input maps of differing
  frequency. The color scale is linear in greyscale from $-1<\alpha<-0.5$, and
  linear in color from $-0.5<\alpha<1$. The gray contours show the shape of
  the source at the lower frequency used to calculate the spectral index map,
  while blue contours trace the source morphology at the higher observing
  frequency used to calculate the spectral index map. These sources were
  randomly chosen from the source list for each category; all show
  matched-resolution data between 2.3\,GHz (gray contours) and 8.4\,GHz (blue
  contours). Contours in these plots are set at 2, 25, and 50 \% of the peak
  flux at each frequency.}
\label{fig:demonstrate}
\end{figure*}

We therefore take special precaution in considering AGN variability and
coverage of the spatial-frequency ($u-v$) plane for the data sets. The
variability exhibited by some radio AGN typically comes in two different
flavours. Both particularly affect compact regions of emission with high
brightness temperature, such as radio cores. The first is intra-day
variability (IDV) due to propagation effects in the interstellar medium. This
signal flickering can occur on sub-day timescales and exhibits a significant
modulation index at frequencies below 5\,GHz, causing fluctuations on average
at the 1-10\% level \citep[e.g.][]{walker98,MASIV}. Intrinsic variability in
radio AGN cores is observed most significantly at frequencies above 5\,GHz,
and occurs on a wide range of timescales. Some radio AGN exhibit slow variations
on timescales of years.
The existence of such variations place limitations on the allowable temporal
spacing between observations at different frequencies. Instantaneous
multifrequency spectra will not be influenced by either effect, while
observations spaced at less than the characteristic timescales of long-term
AGN variability will allay the effects of a wandering high-frequency signal.
For this reason we used only frequency pairs with a time separation of less
than 35 days. Out of the 3982 usable frequency pairs in our data, 3447 were
taken simultaneously, leaving 535 with an average difference in observation
time of about 11 days.




All data processing was done using the {\sc Miriad} software package
\citep{miriad}. For each frequency pair, we inverted the $u-v$ data, tapering
the maxima of $u$ and $v$ to give a circular synthesised beam of the same size
at the two frequencies. A manual inspection of $u-v$ coverage for the pairs
ensured that the sources were sampled enough to provide an accurate
representation of the source morphology at the tapered resolution. Images were
cleaned, restored with the synthesised Gaussian beam of the tapered resolution
to make images of 512x512 pixels, allowing a typical field of view
1.2\,arcseconds to a side (though varying with the resolution of the
observation). An estimate of the image noise level at each frequency was
determined using an off-source region of each image. The number of frequency
pairs we were able to use is detailed in Table \ref{table:nsources}.

The data were imaged in three ways for manual candidate searching. First, for
each resolution-matched frequency pair, the full-field-of-view images at low
and high frequency were plotted in greyscale with logarithmic contours
beginning at three times the noise level of each image. Enhanced plots of the
central image regions were made in a similar fashion, and a spectral index map
of the inner regions was computed, calculating the spectral index
pixel-by-pixel at any point in the image where the flux at both frequency
bands was greater than three times the image noise. Spectral index images were
superimposed with contours for each respective frequency, and the spectral
index value was plotted as a colour image with a sharp desaturation break at
$\alpha = -0.5$ to enhance the appearance of flat-spectrum source components.

We then manually assessed each source, dividing sources into pointlike,
resolved/extended, and multiple flat-spectrum component sources. Representive
sources of each category are shown in Figure \ref{fig:demonstrate}. Sources
with multiple flat-spectrum components were considered preliminary
candidates. In cases where the sources in multi-frequency images appeared
displaced, a manual best-fit alignment was done and the source was
reassessed. We considered a component to be ``flat spectrum'' if its two-point
spectral index was $\alpha\gtrsim -0.6$ at any frequency pair, or if structure
was visible at high frequency which was not detected at low frequency. As
previously noted, sources which contained no detection above a 4$\sigma$ noise
level were removed from the sample.

The nature of preliminary candidates was then assessed by imaging all
available observations of each candidate. Sources were then rejected
from the candidate list if they satisfied any of the following:
\begin{itemize}
	\item[-] In the case of discrete compact flat spectrum components, if suspected components showed spectral steepening at later observation epochs (implying a false detection due to a young, evolving jet component).
	\item[-] If an analysis of the time-dependent movement of suspected flat-spectrum source components revealed the component(s) as members of a steady outward (in some cases, superluminal) flow along a jet axis.
	\item[-] If the full resolution image revealed beam sidelobes as the source of a spurious, flat-spectrum component.
\end{itemize}

Our search returned 12 candidates containing genuine multiple
flat-spectrum components as discussed in more detail below.

\begin{figure*}
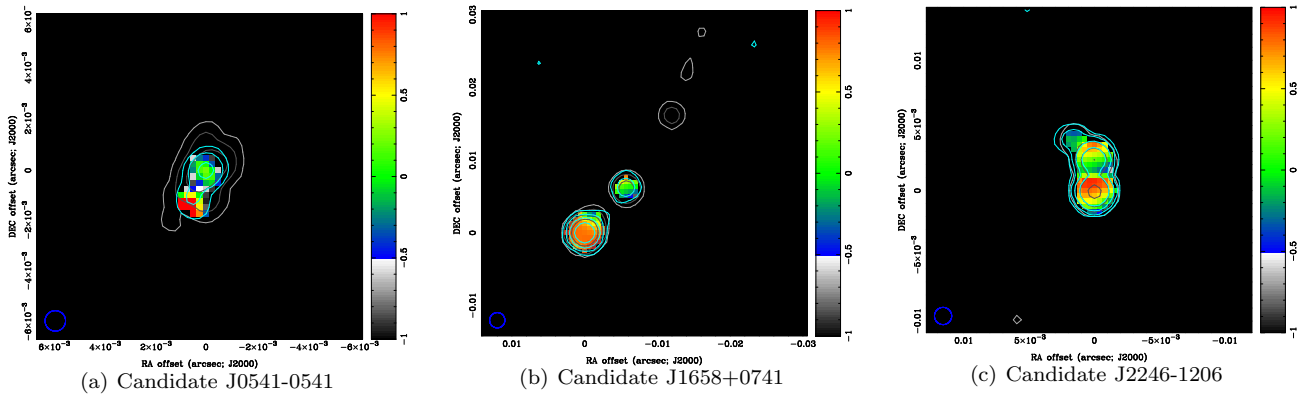

 \subfigure[Candidate J0541-0541]
{
\label{fig:0541}
 \includegraphics[height=0.3\textwidth,angle=270]{ps/paperJ0541-0541.ps}
}
\quad
 \subfigure[Candidate J1658+0741]
{
\label{fig:1658}
 \includegraphics[height=0.3\textwidth,angle=270]{ps/paperJ1658+0741.ps}
}
\quad
 \subfigure[Candidate J2246-1206]
{
\label{fig:2246}
 \includegraphics[height=0.3\textwidth,angle=270]{ps/paperJ2246-1206.ps}
}
\caption{Example two-frequency spectral index and contour maps for three of
  the initial candidates described in \S\ref{sec:indiv}. Coloring of the mapped
  data and contours are as in Fig. \ref{fig:demonstrate}. Contours are set at
  4, 16, 64, and 256 times the RMS image noise drawn from the respective
  frequency's input image.}
\label{fig:cands}
\end{figure*}

\section{Candidates and search results}\label{sec:results}
Of the 3114 sources in our search, 68 percent were found to be pointlike at
our frequency-matched resolution, while the remaining sources were identified
as resolved
or otherwise complex systems with only one flat spectrum component. The number
of sources which exhibited multiple flat-spectrum components comprised less
than 4 percent of the sample, and after the assessment of preliminary
candidates, there remained only 12 candidate binary sources. As an example of
our data quality for candidates, spectral and two-frequency contour images for
three of the candidates below are shown in Figure \ref{fig:cands}.
One candidate was the binary black hole candidate published by
\citet{rodriguezetal06}: 0402+379; an investigation of this source is
available in their publication. We found that the remainder of our candidates
appear to contain regions with active jet injection, which effectively gives
rise to a population of contaminants for the radio spectral index mapping
technique, in some cases requiring further data and analysis to distinguish
such a source from a genuine binary black hole.
Below we use our data and that available from the literature to further
investigate each source and gather evidence for the presence of such
processes, and not a binary black hole, in the sources. 

Ultimately, we report the detection of no new binary systems. We have searched
a sample of considerable size, finding only one spatially resolved binary
black hole out of 3114 radio-bright AGN. Therefore, within the limitations of
our search and data, we find 0.032\% of sources in our sample in which a
binary radio AGN was detected. The implications of this are discussed in later
sections.

\subsection{Candidates with active relativistic injection}\label{sec:indiv}
Eleven of twelve candidates were rejected after the examination of all
available data (in our data, the literature, or available online\footnote{We
made use of the MOJAVE database at http://www.physics.purdue.edu/MOJAVE, which
is maintained by the MOJAVE team \citet{mojave}.}) revealed that the multiple
flat-spectrum components were part of a jet. These sources could be
discriminated from a binary SMBH by proper (superluminal) motion, or by the
resolving of the emission in higher resolution maps and alignment of the
components with a more extensive outflow:
\vspace{3mm}

\noindent \emph{J0111+3906:} 
The dynamical age of the source (370 yr, \citealt{owsianiketal98}, and the
extendedness and alignment of the eastern component along the larger source
structure indicate that it is a young jet.

\noindent \emph{J0541-0541:} 
The southwestern flat spectrum component contains no compact emission at the
high-resolution 8.4\,GHz observation.
This source is shown in Figure \ref{fig:0541}.

\noindent \emph{J0741+3112:}
The persistent total intensity and polarisation observations by \citet{mojave}
clearly demonstrate that this source has a continuous outflow which is
fuelling the flat spectrum hotspot apparent in our maps.

\noindent \emph{J1147+3501:}
\citet{giovannini99} observe superluminal motion of the western flat-spectrum
components.

\noindent \emph{J1223+8040:} 
\citet{pollack03} indicate that there is a jet traversing our flat-spectrum
components.

\noindent \emph{J1347+1217:} 
The multiple flat-spectrum components are all members of an active
superluminal outflow, as shown by the five year modelling of \citet{mojave}.

\noindent \emph{J1459+7140:} 
The southern component is fully resolved at our highest resolutions, and
aligned with a small-scale jet emerging from the northern flat-spectrum
component.

\noindent \emph{J1658+0741:} 
The weaker flat-spectrum component of this source is resolved in our
non-tapered images and is aligned with the northern jet, visible in Figure
\ref{fig:1658}.

\noindent \emph{J1823+7938:}
The resolved structure of the easternmost component in our data
and the proper motion analysis of \citet{britzen}
indicate that it is not a binary radio nucleus. This candidate is shown in
Fig. 2(c).

\noindent \emph{J2246-1206:} 
We conclude from the motion analysis of \citep{mojave} that the northern
component is a jet member. This candidate is displayed in Figure
\ref{fig:2246}.

\noindent \emph{J2253+1608:}
\citet{mojave} show the western flat-spectrum component to be a jet member.

\section{Framework for Interpretation}\label{sec:interp}
The use of radio data as a black hole indicator puts immediate constraints on
both the physical properties and number of sources we will be sensitive to. We
adopt the following expression for the number of sources expected to be found
in our search:
\begin{equation}\label{eq:nexp}
	N_{\rm exp}= \sum_{i=0}^{N_s}f_{\rm bbh}(z_i,m_{\rm{lim}})\cdot P_i
\end{equation}
Here, $N_s$ is the number of objects that have been searched. It is assumed
that a target's redshift $z$ is known (true for 1575 of our objects), and that
it contains at least one detectable black hole of mass $M_\bullet>m_{\rm
lim}$.\footnote{Throughout our analysis we use a value $m_{\rm lim} = 10^8
\,\mbox{M}_\odot$; here we make use of the observed property that radio-bright
quasars have a distinct mass distribution: $\langle {\rm log} (M_\bullet /
\rm{M}_\odot)\rangle = 8.89\pm 0.02$ with a lower cutoff around $10^8\,{\rm
M}_\odot$ \citep[e.\,g.][]{mclurejarvis04}.
We probe only this most massive end of the black hole population, and for a
double detection to be made, the second black hole must exceed a limiting mass
$m_{\rm{lim}}\geq 10^8\rm{M}_\odot$.} $P$ represents the probability that we
would detect a second black hole if the system is a binary, being the
probability that the secondary black hole is both radio-emitting, and bright
enough for a successful detection in our images. This term is explored in
Sec. \ref{sec:RS} below. The fraction of galaxies, $f_{\rm bbh}$, containing a
SMBH binary system at a redshift $z'$ to which we are sensitive will be the
cumulative number of galaxy mergers over the epoch at $z'$ until the epoch at
\mbox{$z''[t_z-t_{\rm vis}]$}, where $t_z$ is the age of the universe at
redshift $z'$, and $t_{\rm vis}$ is the duration that a signature of a double
black hole in the merging system is detectable in our search. Thus, if a
binary stalls at a separation detectable by our technique (bounded by our
observing resolution and the field of view of our images; see
\S\ref{sec:analysis}), galaxies that began to merge when the Universe was the
age of $z''$ will still be occupied by a resolvable binary at $z'$. The
occupation fraction for binaries at a resolvable separation is then given by:
\begin{equation}
	f_{\rm bbh}(z,m_{\rm lim})= \frac{\int_{z'}^{z''}N_{\rm{mrg}}(z,m_{\rm lim})~dz}{N_{\rm{gal}}(z',m_{\rm{lim}})}~,
	\label{eq:fbbh}
\end{equation}
where $N_{\rm mrg}$ is the number of virialized galaxy pairs with SMBHs of
masses above $m_{\rm lim}$ at the given redshift, while $N_{\rm gal}$ is the
total number of galaxies at $z'$ with SMBH masses greater than $m_{\rm lim}$.
The timescale of inspiral over the scales to which our data are senstive
($t_{\rm vis}$) can thus be estimated with a prediction or measurement of the
redshift-dependent merger rate of galaxies containing supermassive black
holes.


Whether we can identify a binary depends on factors explored in the sections
below. The timescale over which we can resolve a binary is determined by our
spatial sensitivity, the black hole masses and binary mass ratio; the relative
time spent at various separations in a binary's inspiral depend on black hole
mass. If the radio luminosity of the black holes somehow depends on the stage
of the merger (for instance if the radio mechanism shuts down as the galaxies
begin interacting, or ignites only after a binary coalescence), an additional
factor needs to be considered in Eq. \ref{eq:nexp}. However, because the
evidence for a relationship between radio fuelling and merger events has until
now been tenuous and estimates for an AGN timescale are so far inconclusive,
we limit this uncertainty by employing two scenarios below that represent the
most optimistic and the most pessimistic expectations for the state of radio
luminosity in the two black holes.

\subsection{Supermassive Black Holes and Radio Emission}\label{sec:RS}
All our targets are radio sources detectable on parsec scales, and it follows
that all contain at least one radio-emitting supermassive black hole. If a
target is a binary system with a resolvable separation between the two black
holes, due to our inspection techniques a second radio-emitting SMBH will have
only been flagged as a candidate if its flux density exceeds four times the
root-mean-squared noise ($\sigma_{\rm rms}$) in its image. This
$4\,\sigma_{\rm rms}$ flux limit corresponds to a luminosity sensitivity limit
of $L_{\rm lim}(\sigma_{\rm rms},z)$. To predict the expected number of binary
AGN to be found in our search and estimate a timescale for binary SMBH
inspiral, we must first quantify the likelihood that the radio emission of a
second black hole will exceed this luminosity limit.

The main unknown contributing to the discussion is how strongly the galactic
and intergalactic environment relate to the production of radio AGN activity
around the central black hole: that is, do the merger and mixing of gaseous,
dusty, and stellar environments contribute strongly the production of radio
activity? Below we assign two scenarios to bound assumptions at either end of
extremes: 1) Pessimistic; the probability that a SMBH paired with a
radio-emitting AGN is itself radio-emitting is no greater than the probability
that a non-binary SMBH at the center of a massive galaxy would be
radio-emitting, and 2) Optimistic; given a black hole that exists in a shared
environment with a radio-emitting AGN, the probability of radio emission in
the second black hole is unity.

Given that the black holes in the progenitors may not encounter significantly
mixed environments until the binary has a relatively small separation, it is
also possible that radio ignition is radially dependent. Regardless, we stress
that for the third stage of inspiral in which theoretically the binary may
stall, the stellar cores around the black holes have merged and the black
holes will have shared many orbits within one another's sphere of influence
and in a common environment.

\subsubsection{Pessimistic scenario}\label{sec:pess}
We can make an estimate of the fewest number of binary sources we would expect
to see by setting $P$ equal to the integrated bivariate luminosity function of
galaxies, $\psi(M,>$$L_{\rm lim})$, where $M$ represents the absolute
optical/infrared magnitude of a galaxy. In this case we assume that mergers
have no influence on the radio-active state of a black hole; that is, that the
probability that the second black hole in a binary would have a detectable
luminosity $L>L_{\rm lim}$ is equivalent to the radio active fraction for
non-merging galaxies selected from a random pool with $M_\bullet\geq m_{\rm
lim}$. We calculate $P$ for each source using a log-linear interpolation of
the integrated bivariate luminosity function of \citet{sadlerandfriends} for
radio AGNs in galaxies with $-24>M_K\geq-25$, taking such galaxies to be
representative of our targets' host galaxies. As roughly 10\% of the
\citet{sadlerandfriends} sample have flat spectra (based on the measurements
of \citet{mason09}), we expect that approximately 10\% of their sources will
be compact down to VLBI scales. To account for this we divide the space
density given in their radio luminosity function by a factor of 10. For our
1575 sources with a known redshift, we find that in this scenario $P$ has a
mean and standard deviation of 0.4\% and 1\%, respectively.

\subsubsection{Optimistic scenario}\label{sec:opt}
In the optimistic scenario, we consider that a SMBH in the presence of a radio
AGN will itself be a radio-loud AGN. This scenario implies that when the two
black holes share a common galactic environment, both black holes will be
radio luminous if the conditions are fit for radio ignition; that is, both
objects will be either radio-loud or not. We take as a given that one black
hole is a detectable radio AGN in each of our images. To determine the
detectability of a second SMBH in such a system (the probability that its
observed luminosity is greater than $L_{\rm lim}$), we would like to avoid the
explicit use of e.\,g. relativistic beaming models, correlations between black
hole mass and radio luminosity, and the levels of flux resolved by our
baselines---the parameters of which are uncertain---by using an estimate
derived from observations in which such effects are inherent.

We therefore take the secondary SMBHs in such systems to have a radio
luminosity distribution equal to the luminosity distribution of solitary,
flat-spectrum radio black holes, which will implicitly include beaming and
mass-correlation effects. The use of the flat-spectrum population assumes that
the emission observed in flat-spectrum radio sources is largely unresolved,
and therefore scalable to the resolutions reached by our VLBI data. We adopt a
probability density function for secondary black hole radio luminosities,
$\rho(z,L)\,dL$, based on the redshift-dependent, flat-spectrum ``pure
luminosity evolution model'' of \citet{dunpea90}. We may then determine the
probability that for a given object in the sample, a second black hole will
have a luminosity above our detection threshold, $L_{\rm lim}$. The
probability that we would have detected the second black hole of a binary AGN
is therefore:
\begin{equation}\label{eq:pdet}
  P = \frac{1}{\rho_n}\int_{L_{\rm lim}}^{\infty} \rho(z,L)\,dL\,.
\end{equation}
The normalisation factor, $\rho_n={\int_{L_{\rm RL}}^{\infty} \rho(z,L)\,dL}$,
sets the range of radio luminosities of radio-loud AGN. We set the lower
luminosity bound of radio AGN at $L_{\rm RL} = 10^{23}$\,W\,Hz$ ^ {-1}$\,sr$ ^
{-1}$, equal to that given by \citet{padovaniRL}. Although the exact value of
$L_{\rm RL}$ debatable, the average $P$ value is not highly sensitive to the
value of $\rho_n$ if $L_{\rm RL}$ is decreased. For images in which the
$L_{\rm lim}<L_{\rm RL}$, the probability of detection is set to one. For the
1575 known-redshift sources in this scenario, $P$ has a mean and standard
deviation of 57.9\% and 23.7\%, respectively.

\begin{figure}
{
 \includegraphics[height=0.48\textwidth,angle=270]{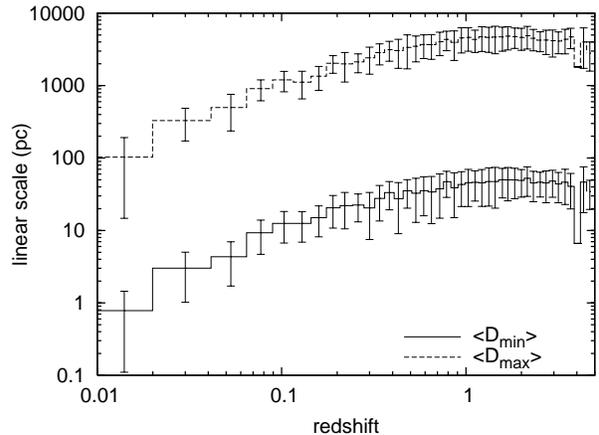}
}
\caption{The upper and lower bounds of the average linear scales searched as a
function of redshift. The error bars represent the standard deviation of
$\langle D_{\rm min}\rangle$ and $\langle D_{\rm max}\rangle$.}
\label{fig:linres}
\end{figure}

\begin{figure}
{
 \includegraphics[height=0.48\textwidth,angle=270]{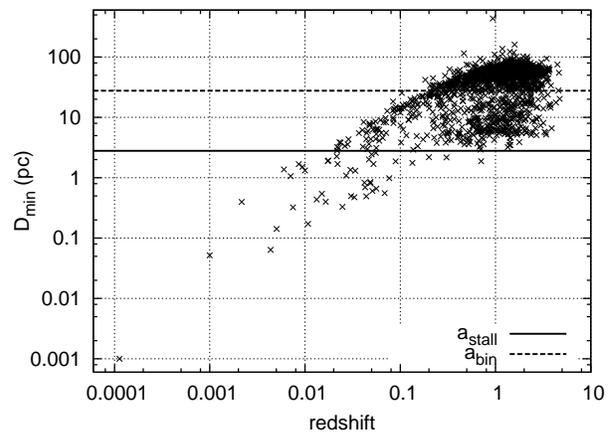}
}
\caption{The minimum resolvable linear scale for each of the sources of known
redshift in our sample. Overplotted are the predicted black hole binary
formation radius and stalling radius for two $10^8\,{\rm M}_\odot$ black
holes. Note that for a binary in a circular orbit, these radii correspond to a
maximum projected separation of twice each value.}
\label{fig:smin}
\end{figure}

\subsection{Host Properties and Limiting Radii}\label{sec:mass}
The predicted binary formation radii and stalling radii mark key points in
possible inspiral rate changes for a post-merger system. In order to determine
what stage of inspiral we have probed with our spatial sensitivity limits, we
require an estimate of the predicted typical $a_{\rm bin}$ and $a_{\rm stall}$
for our sample. As mergers between less massive galaxies are expected to be
more common than high mass mergers, for this study we expect the least massive
(and most common) merger probed in our sample to involve two black holes of
$10^8\,\rm{M}_\odot$. Relevant progenitor galaxy properties can be calculated
to estimate the characteristic scales (e.g. stalling radius, binary formation
radius) that our study probes.  Using two black holes of mass
$10^8\rm{M}_\odot$, each progenitor's velocity dispersion, $\sigma_v\sim
187$\,km\,s$^{-1}$, and bulge mass, $M_{\rm{host}}\sim 6.63\times
10^{10}\rm{M}_\odot$, can be estimated by the $M_\bullet-\sigma_v$ relation of
\citet{ferarrese00} and the $M_\bullet-M_{\rm{bulge}}$ relation of
\citet{haring04}, respectively. If the black holes do not accrete a
significant fraction of their mass during inspiral, the post-merger galaxy
will contain a black hole of $2\times10^8$\,M$_\odot$ and an implied velocity
dispersion of $\sigma_v\sim 216$\,km\,s$^{-1}$. The binary formation radius
for our typical system is thus $a_{\rm bin}=27.7$\,pc, while the stalling
radius is $a_{\rm stall}=2.8$\,pc (equations \ref{eq:abin},
\ref{eq:astall}). In the earliest stage of merger, the virially bound galaxy
pair will have a projected separation $\lesssim 7$\,kpc. We note that while
these radii are the characteristic values for the sources in the sample, more
massive (and less common) mergers between $10^9\,{\rm M}_\odot$ black holes
will have larger radii by a factor of $\sim$10.

\subsection{Spatial Limits of the Search}\label{sec:analysis}
Our technique resolves a large range of spatial scales, over which an inspiral
rate will undergo several changes. To estimate post-merger black hole
evolutionary timescales we must determine the projected spatial separation
range that our data set is sensitive to. For the 1575 sources of known
redshift, the average maximum and minimum projected linear spatial
sensitivities ($\langle D_{\rm max}\rangle, \langle D_{\rm min}\rangle$, both
in units of parsec) as a function of redshift are plotted in Figure
\ref{fig:linres}. The sources are binned in increments matching the redshift
steps of the Millennium Simulation snapshots (see \S \ref{sec:cosmodel}; each
step spans an average of 300\,Myr).

The observed scales are derived as follows: the minimum and maximum linear
projected separation between two objects that we are sensitive to are set by
our angular resolution, $\theta_r$, and half the field of view of our search
maps, given that one AGN will lie at the pointing centre of the observation.
Using the redshift of each source and a flat-universe cosmology with
$H_0=72$\,km\,s$^{-1}$\,Mpc$^{-1}$, $\Omega_m=0.27$, and
$\Omega_\Lambda=0.73$, we calculate $d_z/\theta_z$, the distance (in pc) per
milliarcsecond at that redshift on the plane of the sky. Our smallest resolved
size for that source is then $D_{\rm min} = d_z\theta_r/\theta_z$\,parsecs
($D_{\rm min}$ for individual sources are plotted in Figure
\ref{fig:smin}). Because we searched images of 512 pixels to a side and the
synthesised restoring beam is 3 pixels in breadth, the maximum linear scale we
are sensitive to is:
\begin{equation}
	D_{\rm max} = \frac{512~\theta_r}{6}\cdot\frac{d_z}{\theta_z}~\mbox{parsecs\,.}
\end{equation}
This value represents the largest scale size to which the search of each
source is complete. For sources with known redshift, the sample averages were
$\langle \theta_r\rangle= 6.5$\,mas, $\langle D_{\rm min}\rangle= 40\pm
25$\,pc and $\langle D_{\rm max}\rangle= 3415\pm 2133$\,pc.

The limited spatial separation window over which we have searched each source
for a double AGN and the potential for drastic changes in inspiral rate at
various stages of the binary evolution compel us to divide our measurement of
inspiral rates into three stages. We do so by segregating the sources by their
spatial resolution into groups which represent sensitivity to three stages of
merger evolution: the stage at which the stellar cores have not yet merged
(merger stage 1), the stage leading to binary formation in which the cores
merge and the binary forms (stage 2), and the final stage to which we can
probe spatially, in which the black holes are a binary and evolve to the
predicted stalling radius (stage 3).  In each of these groups, we consider
that a linear resolution of $D_{\rm min}=a_{lo}$ will allow the detection of
sources with an orbital semi-major axis of $a_{lo}$ roughly 80\% of the time,
assuming systems with a random distribution of inclination angles. By setting
the constraint on a group that $D_{\rm min}<a_{lo}$, we ensure that the
searched nuclei in that group will be sensitive to a binaries with axis $a\geq
a_{lo}$ for $>80$\% of the pair's orbit. Each group is statistically sensitive
in this way to pairs with an orbital axis of up to $\langle D_{\rm
max}\rangle$ of the group.

The three groups are divided at \emph{(1)} \mbox{$27.7<D_{\rm
min}<120$\,parsecs,} \emph{(2)} \mbox{$2.8<D_{\rm min}<27.7$\,parsecs}
(sensitivity down to our limiting $a_{\rm bin}$), and \mbox{\emph{(3)} $D_{\rm
min}<2.8$\,parsecs} (giving sensitivity down to the minimum $a_{\rm
stall}$). Because sources with very high resolution observations will not have
a large value of $D_{\rm max}$ and thus not probe large separations, the
source group (3) is not included in (2), and in turn these two groups are not
included in (1).  The divisions give a sensitivity to binaries at separations
over three ranges of scales:
\begin{itemize}
	\item[(1)] $120<a<4757$\,parsecs
	\item[(2)] $27.7<a<2519$\,parsecs
	\item[(3)] $2.8<a<450$\,parsecs
\end{itemize}
There were 1035, 497, and 43 sources in groups one, two, and three, respectively.


\begin{figure}
{
 \includegraphics[height=0.48\textwidth,angle=270]{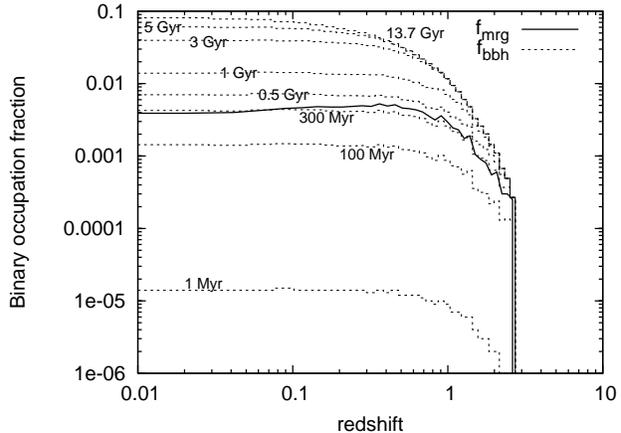}
}
\caption{The redshift-dependent fraction of galaxies containing
binary or double SMBH black hole systems for various inspiral times, based on
the merger rate predictions of \citet{deluciablaizot} for mergers with black
holes of masses $m_1, m_2\geq 10^8\,{\rm M}_\odot$.}
\label{fig:fmrgfbbh}
\end{figure}

\subsection{Galaxy Merger Rates}\label{sec:cosmodel}
The merger rate of galaxies containing SMBHs given a $\Lambda$CDM universe in
which merging is the primary mechanism for galaxy growth can be predicted
using results from the Millennium Simulation \citep{springel05}. The
Millennium simulation is a large-scale N-body simulation which tracked the
evolution of dark matter halos in a co-moving cubic volume 500\,Mpc\,$h^{-1}$
to a side ($h=H_0/100$\,km\,s$^{-1}$\,Mpc$^{-1}$). \citet[][]{deluciablaizot}
applied a semi-analytical prescription to the Millennium simulation to track
the evolution and merger histories of individual galaxies within the
Millennium volume. Their catalogue can be queried through an online
interface.\footnote{See http://www.g-vo.org/Millennium}

Using the Millennium catalogue, we constructed the redshift-dependent
distribution of galaxies and mergers in which each galaxy contains a
supermassive black hole of \mbox{$M_\bullet> 10^8\,\mbox{M}_\odot$}. For
comparison with the black hole binary occupation fraction, we show these
values as a pair fraction, where \mbox{$f_{\rm{\rm mrg}} = N_{\rm mrg}/N_{\rm
    gal}$}; here, $N_{\rm mrg}$ and $N_{\rm gal}$ are the number of mergers
and the total number of galaxies, respectively, with
\mbox{$M_\bullet>10^8\,{\rm M}_\odot$} at the given simulation snapshot. A
``merged'' pair in the Millennium Simulation and the De Lucia \& Blaizot
framework is a pair which has become gravitationally bound in the
$\sim$300\,Myr between two adjacent snapshots.

Figure \ref{fig:fmrgfbbh} shows the redshift-dependent pair fraction for
$m_{\rm{lim}}>10^{8}\rm{M}_\odot$ and the corresponding black hole binary
occupation fraction (computed from eq. \ref{eq:fbbh}) for a range of inspiral
times. The merger distribution peaks at a redshift of about $z=0.7$.

\section{Limits on inspiral timescales}\label{sec:timeresults}
Following Eq.\,\ref{eq:nexp}, for the merger stage groups with no SMBH binary
detections, we place an upper limit on $t_{\rm vis}$ at $N_{\rm exp}=0.5$. The
one source in which a binary black hole was detected is part of the second
merger stage group; for this group/region only are we able to make a
measurement of (rather than put a limit on) $t_{\rm vis}$, with errors
bounding where $t_{\rm vis}$ gives a value of \mbox{$0.5 < N_{\rm exp} <
1.5$}.

\begin{figure}
{
 \includegraphics[height=0.48\textwidth,angle=270]{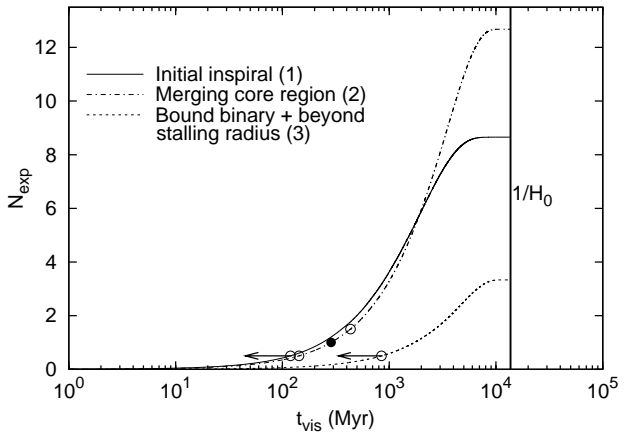}
}
\caption{The cumulative expected number of sources and limits on the inspiral timescale for each spatial-sensitivity limited group using the optimistic scenario (\S\ref{sec:opt}) in which given one radio-active black hole in a binary, the second will also be radio-emitting. For groups 1 and 2, the upper limit set on $t_{\rm vis}$ is marked with an arrow, while for group two the measured value of $t_{\rm vis}$ is marked by a filled circle, with open circles indicating the error range.}
\label{fig:optnexpvtvis}
\end{figure}

The timescale measurements for the optimistic scenario are summarised in
Figure \ref{fig:optnexpvtvis}, which illustrates the number of black holes
expected to be found in each group as a function of $t_{\rm vis}$, and Table
\ref{table:timescales}, which gives the numerical values of our limits. Under
the pessimistic set of assumptions, we would not expect to see any sources,
given that even with $t_{\rm vis}=1/H_0$, we find $N_{\rm exp}=0.025, 0.20,$
and $0.16$ for groups 1, 2, and 3, respectively. The implications of this will
be discussed in the next section.

In Table \ref{table:timescales} we report the timescales measured from both
our 1575 sources of known redshift, and from the full sample with an assumed
redshift distribution. While we do not use the less reliable ``full sample''
timescale values in the analysis below, we stress the limiting power of the
sample if we were to have a full sample with measured redshifts, by assuming
the remaining sources have a redshift distribution equivalent to the other
1575. This has the effect of roughly doubling $N_{\rm exp}$ for each group,
and as such works only to lay more stringent limits on $t_{\rm vis}$.

Finally, although the number of sources in group one was the largest, these sources
also had a tendency to be of higher redshift, with a net effect of fewer
expected detectable binary black holes due to the lower merging fraction of
SMBH hosts at redshifts $z\gtrsim1$. For stalling studies, the observation of
low-redshift sources is therefore doubly beneficial: not only does it allow
smaller scales to be resolved in the sources due to their proximity, but it
affords a higher predicted success rate due to the increased merger rate of
SMBH-hosting galaxies at low redshift.

\begin{table}
\centering
\begin{tabular}{r r r}
       & \multicolumn{1}{c}{\textbf{Known redshift}} & \multicolumn{1}{c}{\textbf{Full sample}} \\
\hline
      \textbf{Group 1}& $<$0.119             & $<$0.060  \\
      \textbf{Group 2}& $0.286^{+148}_{-143}$& 0.143  \\
      \textbf{Group 3}& $<$0.846             & $<$0.416  \\
      \hline
      \textbf{Net}    & $<$1.25\,Gyr         & $<$619\,Myr\\
\end{tabular}
\caption{ A summary of the $t_{\rm vis}$ measurements (in Gyr) placed by our
known redshift sources (1575) and the full sample with an assumed redshift
distribution (3114 total). The values given are for the set of optimistic
radio luminosity assumptions (\S\ref{sec:RS}). Groups 1--3 are as detailed in
\S\ref{sec:analysis}. Although the groups overlap in their linear separation
sensitivities, we quote a ``Net'' timescale that gives a limit on the total
timescale of inspiral from galaxy virialization to SMBH separations beyond the
stalling radius.
}\label{table:timescales}
\end{table}

\section{Discussion}\label{sec:discussion}
The question we aim to answer with this search is whether there is
observational evidence for stalled supermassive binary black hole systems:
that is, what is the evolution timescale down to the stalling radius for a
post-merger supermassive black hole pair? For our methods, this question is
intertwined with how closely correlated the radio emission is between two
black holes in a pair. Our sample allows us to address this correlation and
inspiral timescales over three stages of post-merger SMBH pair evolution. This
is discussed below, followed by a brief consideration of the consequences of
our results for the detection of gravitational waves from SMBH binary systems
by pulsar timing arrays.

\subsection{Post-merger efficiency, radio ignition, and SMBH binary stalling}
The timescales measured for our three merger stages represent the time a
pair's orbital separation takes to evolve through the spatial sensitivity
window of each stage.
For our merger stage groups one and two ($\sim$virialisation down to
SMBH binary formation), it is expected that
purely dynamical friction will act on the black holes.
Formally, the dynamical friction timescale in equation \ref{eq:tdf} over the
range probed by our first and second merger stage groups is $\sim$3.7\,Gyr and
900\,Myr, respectively, for an object of mass $10^8\,{\rm M}_\odot$.

For merger stage group one, the optimistic scenario timescale limit is smaller
than the dynamical friction timescale by a factor of $\sim$30.
From this disagreement, it follows that one of the following must be true: 1)
the relative radio state of two SMBHs in merging galaxies is not strongly
correlated before the hosts' stellar cores have merged, and/or 2) The
centralisation of supermassive objects in a post-merger system is highly
efficient (in excess of dynamical friction).

Considering merger stage group two, in which we have detected one binary SMBH,
the timescale under the optimistic set of assumptions gives a value more
consistent (a factor of $\sim$2 within the error bounds) with dynamical
friction. Because in the pessimistic limit we would not expect to see any
sources, it is implicit that the pessimistic assumptions are not strictly
valid for this stage of merger. This is furthermore indicative of some radial
dependence of the relative radio state of two SMBHs in a post-merger
environment, such that the SMBHs have a correlated radio active state at later
stages of SMBH inspiral, after the pair has shared a common galactic
environment for many orbits.


We have set the cutoff for the third merger stage group to be sensitive to the
theoretical stalling radius of \citet{merritt06}. While the limits that we can
place on this stage of merger are the least stringent due to the small source
count in the limiting group, they represent the first observational
exploration of stalling. If the evidence exhibited by merger stage group 2
holds---that the relative radio state between two black holes is more strongly
correlated at late stages of binary evolution---our results are in agreement
with a progression through the stalling radius in much less than a Hubble
time; that is, we find no evidence for systematic stalling of supermassive
binary black hole systems.
These results, however, exist with the significant caveat: if the most
pessimistic assumptions are true, we would not expect to see any paired SMBHs
regardless of the length of post-merger inspiral.

As previously noted and reflected in table 2, we have only used the
information from roughly half of our sources to acquire timescale
measurements. The lack of detections in the additional 1539 sources suggests
that the measured limits will only be more stringent, however would be highly
dependent on the redshift distribution of the remaining sources.

\subsection{Consequences: gravitational wave background and pulsar timing arrays}
The measured inspiral timescales and lack of evidence for stalled supermassive
systems are pertinent results for pulsar timing arrays, which are sensitive to
the most massive of binary mergers such as those explored by this study. While
we have primarily accentuated the brevity of timescale to pass through the
stalling stage, we note also that the timescales measured for merger group two
(in concert with time spent at other separations) are sufficiently large that
they may have non-negligible effects on predictions of the gravitational wave
background signature in the local Universe from SMBH binaries. Typically,
estimates of expected binary black hole contributions assume efficient or
instantaneous inspiral into the gravitational wave regime
\citep[e.g.][]{sesanagwb}, using merger rates as binary coalescence rates. The
inspiral times measured here may cause the galaxies merging at a redshift of
$0.7$, at which the merger distribution peaks, to emit gravitational radiation
at a much later epoch, causing a larger expected amplitude for the
astrophysical gravitational wave background than previously estimated. In
addition, it is likely that this effect would cause an increase in the
predicted number of supermassive binary sources whose gravitational radiation
is strong enough to be resolved from the gravitational wave background by
pulsar timing arrays \citep{sesanasingle}.

\section{Summary}\label{sec:summary}
We have performed a search of 3114 active galactic nuclei for the presence of
double SMBH systems using a multi-frequency radio imaging technique.
Of the sources searched, only 0402+379
was apparent with our method as a
binary AGN. While this source has already been put forward as a binary black
hole in \citet{rodriguezetal06}, this search represents a significant enough
statistical sample with which to interpret the existence of this source in the
broader cosmological context of binary supermassive black holes.

This searched has probed estimates of the inspiral timescale for binary black
holes of $m_1, m_2\geq10^8\,{\rm M}_\odot$. For the 1575 sources that had
redshift information with which to discern the linear scales probed in the
host galaxy, we obtained timescale estimates for the inspiral of binary
systems in three ranges of binary separation. We have demonstrated
observational evidence against stalled binary black hole systems by
demonstrating that: 1) at late stages of SMBH pair inspiral, the two black
holes are more likely to be in a similarly radio-active state, and 2) SMBH
binaries proceed from separations of $a\simeq 500$\,parsecs to within the
stalling radius estimates of \citet{merritt06} in less than
$\sim$0.5\,Gyr. The implications of this are that supermassive binary systems
do not stall indefinitely at such radii, suggesting there is a yet
undetermined mechanism by which the black holes are able to dispense energy to
the surrounding environment and proceed to coalescence.



The results of this study come with the caveat that they stem on very
small-number statistics; that is, we detected only one binary AGN in the
analysis, and due to unmeasured redshifts, we were unable to use roughly half
of the sources searched in the statistical analysis. Definitive studies of
this type for large numbers of AGN will become possible with the large
collecting area, dense instantaneous $u-v$ coverage, and long baselines of the
planned Square Kilometer Array.

\section{Acknowledgements}
This publication is part of the doctoral thesis of SBS. The author would like
to acknowledge the support of her supervisors M. Bailes, and particularly
R.~N.~Manchester for their assistance in revising and preparing this
manuscript. The author thanks M. Spolaor for his continued support, and
Prof. I. Browne for his insightful feedback on this manuscript.  SBS
acknowledges the Astronomical Society of Australia for financial travel
assistance that contributed to the completion of this paper.



\end{document}